\documentclass[runningheads]{llncs}

\usepackage{amsmath,amsfonts,amssymb,amsthm,amstext,enumerate}
\usepackage{chessfss}
\usepackage{microtype}
\usepackage{epigraph}
\usepackage{fontawesome}
\usepackage[titles]{tocloft}
\usepackage{subfig}
\usepackage{algorithmic}
\usepackage{fancyhdr}

\usepackage{float}
\floatstyle{plain}
\restylefloat{figure}
\usepackage{tikz}
\usetikzlibrary{arrows.meta,decorations.pathmorphing,decorations.footprints,fadings,calc,trees,mindmap,shadows,decorations.text,patterns,positioning,shapes,matrix,fit,
shapes.misc}
\newcommand*\circled[1]{\tikz[baseline=(char.base)]{
            \node[shape=circle,draw,inner sep=2pt] (char) {#1};}}
\usepackage{circuitikz}
\usepackage{caption}
\usepackage{url}
\usepackage{graphicx}
\usepackage{caption}
\usepackage{url}
\usepackage{hyperref}
\usepackage{cleveref}
\let\orgautoref\autoref
\renewcommand{\autoref}
        {\def\equationautorefname{Equation}%
         \def\figureautorefname{Figure}%
         \def\subfigureautorefname{Figure}%
         \def\Itemautorefname{Item}%
         \def\tableautorefname{Table}%
         \def\algorithmautorefname{Algorithm}%
         \def\sectionautorefname{Section}%
         \def\subsectionautorefname{Section}%
         \def\subsubsectionautorefname{Section}%
         \def\chapterautorefname{Chapter}%
         \def\partautorefname{Part}%
         \def\goalautorefname{Goal}%
         \def\reqautorefname{Req.}%
         \def\adviceautorefname{Rule}%
         \def\parameterautorefname{Param.}%
         \def\definitionautorefname{Definition}%
         \def\theoremautorefname{Theorem}%
         \orgautoref}
\usepackage{hyperref}

\newcommand{\tmpx}{}

\fancypagestyle{sec}{\lhead{\tmpx}}
\begin{document}
\title{RADIS: Remote Attestation of Distributed IoT Services
\thanks{This is a revised and extended version of the work previously reported in the 6th IEEE International Conference on Software Defined Systems (SDS-2019). It is published for personal use by the close community of fellow researchers.
You can cite this work as: \textit{Conti, M., Dushku, E., Mancini, L.V.: RADIS: Remote Attestation of Distributed IoT Services. In: 6th IEEE International Conference on Software Defined Systems, SDS 2019. pp. 25–32 (2019)}
}}
\author{Mauro Conti\inst{1} \and
Edlira Dushku\inst{2} \and
Luigi Vincenzo Mancini\inst{2}}
\institute{Department of Mathematics, University of Padua, Padua, Italy \\
\email{conti@math.unipd.it}\\
 \and
Dipartimento di Informatica, Sapienza University of Rome, Rome, Italy\\
\email{\{dushku,mancini\}@di.uniroma1.it}}
\maketitle              % typeset the header of the contribution
\begin{abstract}
Remote attestation is a security technique through which a remote
trusted party (i.e., Verifier) checks the trustworthiness of a potentially untrusted device (i.e., Prover). In the Internet of Things (IoT) systems, 
the existing remote attestation protocols propose various approaches to detect the modified
software and physical tampering attacks.
However, in an interoperable IoT system, in which IoT devices interact autonomously among themselves, an additional problem arises: a compromised IoT service can influence the genuine operation of other invoked service, without changing the software of the latter.
In this paper, we propose a protocol for Remote Attestation of Distributed IoT Services (RADIS), which verifies the trustworthiness of distributed IoT services. Instead of attesting the complete memory content of the entire interoperable IoT devices, RADIS attests only the services involved in performing a certain functionality. 
RADIS relies on a control-flow attestation technique to detect IoT services that perform an unexpected operation due to their interactions with a malicious remote service.
Our experiments show the effectiveness of our protocol in validating the integrity status of a distributed IoT service.

\end{abstract}
\section{Introduction}

The enormous expansion of the Internet of Things (IoT) devices induces the necessity of interoperable IoT systems. The IoT interoperability will allow heterogeneous IoT devices to interoperate and ultimately to support the deployment of large-scale IoT applications. However, due to the limited capabilities of the IoT devices to adopt complex security techniques, IoT systems are increasingly exposed to a huge number of potential attacks  \cite{mirai,Ronen2017}. Hence, a security mechanism that guarantees secure interoperation between devices plays a key role in establishing trust in an interoperable IoT system.

Remote attestation is used as a security protocol that provides reliable evidence about the trustworthiness of an untrusted device.
Typically, the internal state of resource-constrained devices comprises the program binaries stored in the program memory of the device and the run-time state of the software stored in data memory. During a software execution, the content of the program memory remains static, whereas the data memory's contents always change. Most of the existing remote attestation protocols attest only the program memory, thus leaving undetected the run-time attacks, which target the data memory and do not modify the program memory of a device. For instance, a code-reuse attacker may exploit the Return-Oriented Programming (ROP) technique \cite{Shacham2007} to change at run-time the control-flow of genuine sequences of code (i.e., gadgets) loaded on the device's memory and, consequently, produce a malicious code execution. Other run-time attacks do not change the control-flow of a software, but only the data of the software by manipulating the data pointer through Data-Oriented Programming (DOP) \cite{DOP} technique. As the run-time attacks can become pervasive in IoT systems, some recent remote attestation approaches \cite{Kil2009,Abera2016,Zeitouni2017} have been proposed in the literature to check the integrity of the data memory. However, the existing run-time remote attestation schemes can perform attestation only on single devices. Additional research works \cite{Asokan2015,Ambrosin2016,Kohnhauser:2018,PADS}, which have proposed efficient protocols that run attestation over a large number of devices, do not consider the communication data exchanged among devices.

In this paper, we focus on distributed IoT services, and we show that due to the communication data exchanged between services, a compromised service can affect the integrity of the other legitimate invoked services that interact with the compromised one. In particular, a compromised service may maliciously deviate the control-flow of the legitimate invoked services towards a valid but non-authorized state. The naive approach of running a control-flow attestation protocol for each service would not detect such control-flow deviation because the software of the invoked service is genuine and the deviation is caused due to the corrupted input received. To this end, our work considers interoperable IoT devices and aims to check the integrity of distributed IoT services that run on these devices. 
We propose a remote attestation protocol that detects the control-flow deviation of legitimate services, which is affected by an adversary who has not directly compromised this service but has compromised another service that interacts with the former.

\textbf{Our Contribution:} The contributions of this paper are threefold:
\begin{itemize}
\item We highlight the need for the attestation of distributed IoT services by demonstrating that a compromised service in a distributed IoT service can induce malicious behavior on genuine services. 
\item We define the required security properties for distributed IoT services and describe the adversary model.
\item We present RADIS, a remote attestation protocol for distributed IoT services and provide the performance evaluation.
\end{itemize}

\textbf{Outline:} The remainder of the paper is organized as follows. In \autoref{sec:related_works}, we provide an overview of the current state-of-the-art remote attestation approaches and provide a comparison with our work. We describe the problem in \autoref{sec:radis_problem_setting}. In \autoref{section:radis_system_model}, we present the system model, and in \autoref{section:radis_adversary_model}, we describe the adversary model and the required security properties. We introduce the preliminary concepts in \autoref{section:radis_preliminaries} and provide our protocol details in \autoref{section:radis_solution_description}. The evaluation of protocol is shown in \autoref{section:radis_evaluation} and security analysis in \autoref{section:radis_security_analysis}. Finally, the paper concludes in \autoref{section:radis_conclusions}.

\section{Related Works}
\label{sec:related_works}

In this section, we provide an overview of previous work related to remote attestation protocols. We focus particularly on attestation of distributed services in traditional systems and state-of-the-art remote attestation protocols in IoT systems.

\textbf{Attestation of distributed services.}
Shi et al. \cite{Shi} propose BIND as a fine-grained attestation scheme for traditional distributed systems. At the attestation time, BIND attests only some selected piece of code for each service, following the assumption that programmers annotate the most critical parts of the service software. BIND measures a critical code immediately before entering in the code execution and uses a sand-boxing mechanism to serve as a protection for ensuring the untampered code execution. However, BIND does not address attacks that happen in the intermediary code that is not predefined by programmers for attestation. In our protocol, the runtime attestation takes into consideration the entire sequence of the distributed services without limiting the attestation only to a predefined section of code.
The work presented by Gu et al. \cite{Gu2009} propose an integrity attestation that aims to check whether a subroutine of a program is executed correctly. The proposed scheme leverages Trusted Computing Group (TCG) attestation to build a trust chain rooted at Trusted Platform Module (TPM) for function execution. The attestation schemes presented in \cite{Shi} and \cite{Gu2009} are not designed for resource-constrained devices.

\textbf{Collective attestation.}
In IoT systems, collective attestation schemes aim to verify the internal state of a large group of devices in a more efficient way than attesting each of devices individually. Asokan et al. \cite{Asokan2015} propose SEDA as an attestation approach which constructs the interconnected network as a spanning-tree. In this scheme, each device statically attests its children and reports back to its parent the number of children that successfully passed the attestation protocol. In the end, an aggregated report with the total number of the devices successfully attested will be transmitted to the Verifier. The weakest point of this protocol is that a compromised node can impact the integrity of the attestation result of all its children nodes in the aggregation tree.
This problem was later tackled in the work presented by Ambrosin et al. \cite{Ambrosin2016}. There, the authors propose a scalable attestation protocol with untrusted aggregators (SANA) which relies on the use of a multi-signature scheme. In SANA, devices sign the attestation responses and an aggregation of the signatures is used to validate the network in a constant time.
The basic assumption followed by both SEDA and SANA is that the network is fully interconnected. Recently, Kohnh\"{a}user et al. \cite{Kohnhauser:2018} and Ambrosin et al. \cite{PADS} rule out this assumption and propose an efficient protocol for highly dynamic networks. In these proposals, each device performs the local attestation at the same point in time and shares the individual result with other devices in the network. Then, devices use the consensus algorithm to gain knowledge about the state of the other devices in the network. At the attestation time, the verifier can perform the attestation over a random device, which will report the consensus state of the entire network. Additionally, Ibrahim et al. \cite{Ibrahim:2016} and Kohnh\"{a}user et al. \cite{Kohnhauser:2017} have presented collective attestation schemes that are able to detect invasive physical attacks by following the assumption that an adversary needs to shut a device down for a non-negligible amount of time in order to physically tamper the device.\\
However, the collective attestation schemes do not consider the flow of the interactions between devices and the communication data that go from one device to another. Therefore, these schemes detect devices that are running a modified software, but they do not check whether the devices with legitimate software are executing a task on malicious data. We argue that, in a distributed system, a service victim of a run-time attack can propagate malicious behaviour to all the devices that requested that service, even though the software running on those devices is legitimate.
Additionally, the existing collective attestation schemes verify only the integrity of the static program memory without providing a validation mechanism for the data memory. Thus, runtime attacks remain undetected.

\textbf{Dynamic attestation.}
Dynamic attestation approaches aim to verify the run-time state of the Prover during the normal software execution.
Kil et al. \cite{Kil2009} propose ReDAS as an attestation protocol that verifies the properties of the run-time behaviour of the Prover. When any of the properties is violated, ReDAS stores the relevant evidence in a Trusted Platform Module (TPM). ReDAS checks the system integrity only at system calls, and it traces only the order of the launched modules in a system. Therefore, it does not detect the malware presence between system calls, and it does not check the runtime flow of the instructions of a specific module.
%The existing remote attestation protocols focus on ensuring the integrity and authenticity of software running on devices. 
Abera et al. \cite{Abera2016} propose  C-FLAT, a complete attestation of the run-time state of the Prover. During the execution, each software instruction is reported into a so-called ``trusted anchor'' and from there, a hash engine mechanism accumulates the sequence of the instructions into a single hash value that represents the entire control flow of the Prover's state. A Verifier, who has initially computed and stored a set of all the possible valid hashes of the Prover, can detect control-flow attacks since a Prover targeted with a control-flow run-time attack will report an unexpected hash value to the Verifier. This work is extended by Dessouky et al. \cite{Dessouky2017} which present a practical version of C-FLAT named LO-FAT. Instead of the software instrumentation used in C-FLAT, LO-FAT explores the features of the microcontroller to intercept the instructions, providing in this way an implementation of C-FLAT with low overhead. LiteHAX \cite{Dessouky:2018} extended C-FLAT by proposing a hardware-based protocol that detects both control-flow and data-flow attacks for a specific architecture of embedded devices.
ATRIUM \cite{Zeitouni2017} proposes a hardware-based runtime attestation protocol that is resilient against Time of Check Time of Use attacks. ATRIUM attests both executed instructions and the control-flow.

However, the existing dynamic attestation protocols follow the single-device-attestation approach and do not provide a complete evidence of the integrity of the distributed IoT services.

\section{Problem Description}
\label{sec:radis_problem_setting}

\begin{figure}[h!t]
\centering
\includegraphics[scale=0.9]{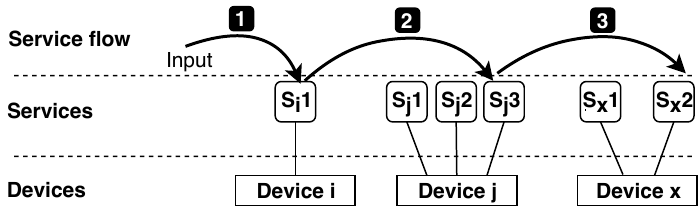}
\caption{Service Flow of IoT devices}
\label{fig:service_flow}
\end{figure}

We consider an interoperable IoT system as shown in \autoref{fig:service_flow}, where different IoT devices provide a set of services that interact together to perform a task. 
The sequence of all the services involved in performing a task is called
Service Flow, and the notation for the Service Flow depicted in \autoref{fig:service_flow} is $S_i1 \rightarrow S_j3 \rightarrow S_x2$. The set of services $S_i1 \rightarrow S_j3 \rightarrow S_x2$ communicating with each other to support the operation forms also a distributed service. Note that a given distributed service can follow a different service flow based on different invocations, depending, for example, on the input parameters.

\begin{figure}[ht]
\centerline{\includegraphics[scale=0.85]{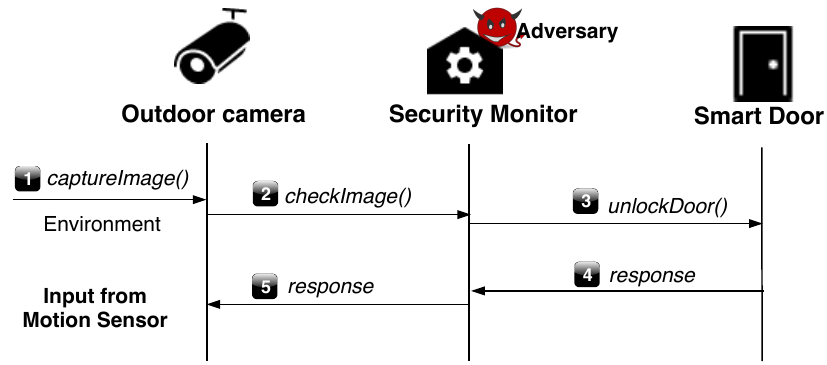}}
\caption{Device interaction in Smart Home IoT System}
\label{fig:overview}
\end{figure}

As a motivating example, we consider a Smart Home IoT system enabled by the interoperation between services of three IoT devices: an Outdoor Camera, a central Security Monitor, and a Smart Door. A motion sensing Outdoor Camera observes outside the main door of the home, and when any movement of objects or people is detected, the camera captures an image and reports it to a Security Monitor. Once the Security Monitor gets the captured image, it analyzes the image, and if it identifies a family member, it sends an unlock command to open the Smart Door, as shown in \autoref{fig:overview}.  The service flow in this scenario is:  $captureImage()$ $\rightarrow$ $checkImage()$ $\rightarrow$ $unlockDoor()$.

Devil's Ivy attack \cite{senrio} shows how an attacker exploits a vulnerability in a widely used library to take control over a security camera. Here, the attacker uses Return-Oriented Programming (ROP) technique \cite{Shacham2007} to change at runtime the execution flow of genuine pieces of code loaded on the device's memory, and consequently, produce a malicious code. As these attacks can become pervasive in IoT systems, a prominent requirement for the attestation schemes is the detection of run-time attacks, which target the data memory and do not modify the program memory of a device. The attestation of the data memory of individual devices requires the execution of a single-device control-flow attestation protocol that detects subverted control flows. One possible example of such attestation protocol is C-FLAT \cite{Abera2016}. In the case the device is not compromised, a control-flow attestation protocol, running on a single device, will report the benign state of the device. For instance, when a single-device control-flow attestation protocol attests a genuine Smart Door, it will ensure its correctness. 

Now, consider an adversary that attacks another device of the distributed service, e.g., the Security Monitor device. In particular, the adversary can corrupt the security monitor's data pointers at run-time or modify the communication data yielded by the Security Monitor. After this attack, a single-device control-flow attestation procedure executed on the Smart Door will report again the correctness of the Smart Door. This is because the adversary has not changed the software of Smart Door and has not deviated its control-flow.
However, even though the adversary is located only in the Security Monitor and the Smart Door passes all the checks of the control-flow attestation procedure, we show that the Smart Door can be forced into an incorrect state.

By compromising the Security Monitor device, the adversary is able to generate malicious software executions on the Security Monitor that can produce malicious data and can influence the current behavior of the other interconnected devices. As a consequence, the state of the Smart Door may be corrupted by the commands invoked maliciously from the Security Monitor to the Smart Door. For example, an $unlockDoor()$ command initiated as result of an attack in Security Monitor can open the door even if the camera has not captured the image of a family member. We thus argue that the Smart Door may have a genuine software, but its behavior is not legitimate if it is performing an unexpected operation due to the command or input that it received from a malicious code executed in the Security Monitor device.

To detect this attack, one could think to run a single-device control-flow attestation protocol on every device of the IoT system. Indeed, the control-flow attestation protocols, running on each individual IoT device, will detect the devices which contain corrupted control-flow information on their data memory. Since the adversary has modified only the value of one variable on the Security Monitor and has not performed any control-flow attack to any device, the control-flow attestation protocols, running on each of the three devices of our scenario, would report all the devices in a legitimate state. Hence, the control-flow deviation of the Smart Door remains undetected.

\begin{figure}[ht]
\centerline{\includegraphics[scale=0.9]{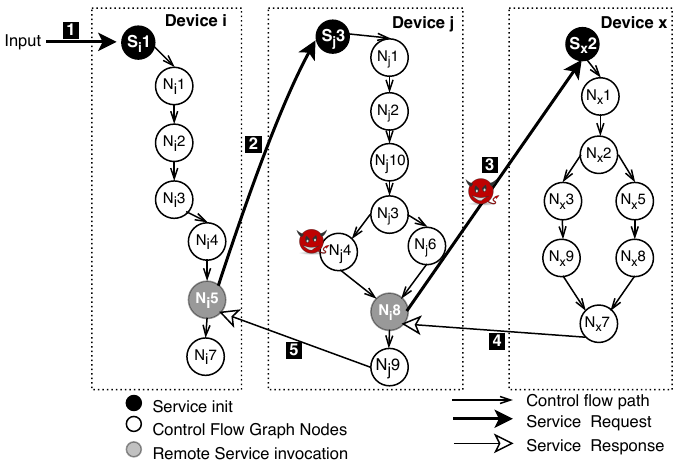}}
\caption{Control flow of the distributed service in \autoref{fig:overview}}
\label{fig:control_flow}
\end{figure}

To clarify the effect of an attack on a distributed service, in \autoref{fig:service_sourcecode} we illustrate the pseudo-code of the three services involved in the aforementioned service flow: $captureImage()$ $\rightarrow$ $checkImage()$ $\rightarrow$ $unlockDoor()$. Based on the instructions of this pseudo-code, for each service is constructed a Control Flow Graph, where each node of the graph presents an instruction, as shown in  \autoref{fig:control_flow}. During the usual operation, each service follows the intended control-flow and then invokes a service call to the next device.

\begin{figure*}
\centering
\begin{tabular}{|p{0.33\linewidth}|p{0.33\linewidth}|p{0.33\linewidth}|}
\hline
\scriptsize{\textbf{Device $i$: Outdoor Camera}} & \scriptsize{\textbf{Device $j$: Security Monitor}} & \scriptsize{\textbf{Device $x$: Smart door}} \\ \hline
\noindent
\begin{minipage}{0.33\textwidth}
\begin{algorithmic}[1]
\tiny
     \algsetup{linenosize=\tiny}
  	 \scriptsize
  	 \STATE \textbf{captureImage()} \{
     \STATE $motion \gets sensor.value();$
     \IF{$motion$}
     \STATE $img \gets~~~~~~~~\break camera.capture();$
     \STATE \underline {$checkImage(img);$}
     \ENDIF
     \STATE \}
     \label{outdoor_camera_code}
     \end{algorithmic} 
\end{minipage}
&
\begin{minipage}{0.33\textwidth}
\begin{algorithmic}[1]
\tiny
     \algsetup{linenosize=\tiny}
     \scriptsize
     \STATE \textbf{checkImage(img)} \{
     %\STATE $cmd \gets false;$
     \STATE $member \gets~~~~~~~~\break searchFamily(img);$
     \IF{$member$ is false}
       % \STATE $verifyUnknownImage(img);$
       %\STATE $notify();$
       \STATE $cmd \gets false;$
     \ELSE 
     	\STATE $cmd \gets true;$
     	%\STATE \underline{$unlockDoor(cmd);$}  	
     \ENDIF
     \STATE \underline{$unlockDoor(cmd);$} 
     \STATE \}
     \STATE service: $searchFamily()$
    % \STATE service: $verifyUnknownImage()$
     %\STATE service: $notify()$
     \label{security_monitor_code} 
\end{algorithmic}
\end{minipage}
&
\begin{minipage}{0.33\textwidth}
\begin{algorithmic}[1]
\tiny
     \algsetup{linenosize=\tiny}
     \scriptsize
     \STATE \textbf{unlockDoor(cmd)} \{
     \IF{$cmd$ is true}
     \STATE $unlock();$
     \ELSE \STATE $lock();$
     \ENDIF
     \STATE \}
     \STATE service: $lock()$
     \STATE service: $unlock()$
     \label{smart_door_code}
\end{algorithmic}
\end{minipage} \\ \hline
%$(a)$ \scriptsize{$service$ $S_i1$: \textit{captureImage()}} & $(b)$ \scriptsize{$service$ $S_j3$: \textit{checkImage()}} & $(c)$ \scriptsize{$service$ $S_x2$: \textit{unlockDoor()}} \\ \hline

\end{tabular}
\caption{Pseudo-code of the service flow in \autoref{fig:overview}}
\label{fig:service_sourcecode}
\end{figure*}
The adversary located in Security Monitor ($Device$ $j$) performs an attack in $N_j4$ to maliciously assign the variable $cmd$ with the value ``$true$''. The service execution will then proceed to Node $N_j8$ to call the service $unlockDoor\left(cmd\right)$, as shown in  \autoref{fig:control_flow}. Note that when the execution flow reached at Node $N_j4$, the variable $cmd$ was assigned as ``$false$''. 
The compromised argument $cmd$, produced by the adversary in Security Monitor, is used in node $N_x2$ of Smart Door ($Device$ $x$) as a decision-making variable that defines the further operations of Smart Door. This means that Smart Door, even though is running a genuine software, can maliciously run $unlock\left(\right)$ command in Node $N_x3$ because of the compromised argument received ($N_x2$ goes into $N_x3$ instead of going into $N_x5$).

Consider now another type of adversary that does not change the software of the services, but modifies the communication data between Security Monitor and the Smart Door. For instance, an adversary that is able to carry out a man-in-the-middle attack in the network can modify the data between node $N_j8$ and $S_x2$ to set $cmd$ as ``$true$''.  Such adversary will still be able to deviate the control-flow of the Smart Door even though the software of Security Monitor and software of the Smart Door are both genuine. 
The attacks described above show that a compromised device ($Device$ $j$) induces a malicious control-flow deviation into a subset of IoT devices, even though the software running on the subset of the devices is not altered in any way by the attacker. Therefore, to produce a correct attestation response of a distributed service, the attestation protocol should not only detect the compromised services, but also the services that are performing a non-intended operation due to their interactions with the infected service.

\textbf{Note.} The goal of our work is to verify whether a distributed services is performing an intended operation and we do not intend to check the integrity of the entire data processed by each service. Considering that some data attacks can have only an isolated impact on the overall operation of a distributed service, our protocol does not consider the data attacks which impact neither the control-flow of an individual service nor the control-flow of the invoked services.

\section{System model}
\label{section:radis_system_model}
We consider a distributed IoT system, where each heterogeneous IoT device $D_i$ provides a number $n_i$ of services. %each uniquely identified as $S_iu$ where $1 \leq u \leq n_i$. 
In a typical distributed IoT service, each service invokes an explicit service request to another service according to a predefined interaction model. A distributed IoT service may follow various service flows at run-time,, thus, the aim of the attestation mechanism is to check the integrity of a distributed service by verifying that a given service flow is legitimate.
In modelling the attestation scheme of a distributed IoT service, we consider the presence of the following entities: 

\begin{itemize}
\item Device $D_i$: a number of interconnected devices that compose a distributed IoT system. Each device hosts $n_i$ different services, each uniquely identified as $S_iu$, for $1 \leq u \leq n_i$.
%Each device provides a set of services, which do not run concurrently in the device. Each service can be identified by a unique name in the distributed system.
\item System operator $OP$: responsible for the trusted deployment of the distributed IoT system.
\item Verifier $Vrf$: a trusted external party who checks the integrity of a service flow of the distributed IoT system. $Vrf$ may be different from $OP$. $Vrf$ has access to the binaries of all the services deployed on the distributed IoT system. The attestation runs periodically at an arbitrary time determined by $Vrf$.
\end{itemize}

\begin{figure}[ht]
\centerline{\includegraphics[scale=0.54]{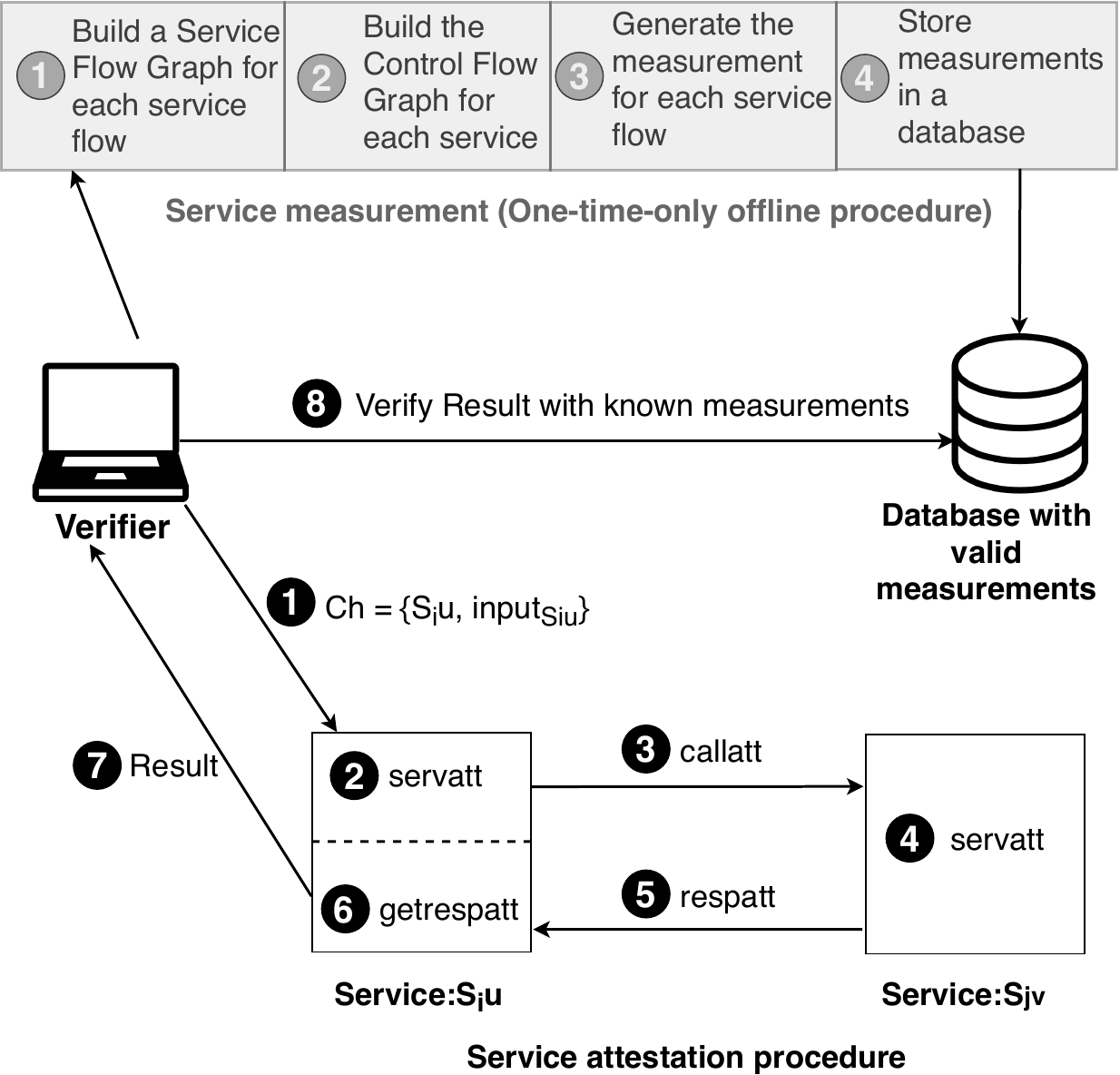}}
\caption{System model of remote attestation of a distributed IoT service, which consists of two services $S_iu$ and $S_jv$.}
\label{fig:system_model}
\end{figure}

Initially, an IoT system operator $OP$ validates the identities of the devices, authorizes their access, and verifies the correct version of the software and services available on them. Then, a Verifier $Vrf$, responsible for the integrity check of the distributed services, performs an offline procedure to measure all genuine services that compose the distributed IoT system. During the service measurement procedure, $Vrf$ considers the legitimate service flows and all possible legitimate control-flows of the genuine services that compose a service flow. Next, $Vrf$ generates the measurement for each service flow, and at the end of this procedure, $Vrf$ stores in a database a single hash value for each legitimate service flow. A conceptual overview of our system model is depicted in \autoref{fig:system_model}.

At the attestation time, $Vrf$ sends an attestation request \circled{\small 1} to the device hosting the first service of a given service flow. 
Upon receiving the attestation request, the device initiates the attestation process for the intended service \circled{\small 2}. During the execution of the service, a run-time trace module traces all the instructions of the services and invokes a hash module to compute an accumulative hash for the entire control-flow path that the service follows at run-time. Then, each service invocation comprises also the attestation result. This process binds all the services attestation reports generated through the entire service flow \circled{\small 2} - \circled{\small 6}. After completion, the first service of the service flow sends to $Vrf$ the final attestation report of the entire service flow \circled{\small 7}. In the end, $Vrf$ validates the received result with the known measurements stored previously in the database \circled{\small 8}. If the final attestation result matches with one of the pre-calculated values, $Vrf$ ensures that the distributed service is in the legitimate state. Otherwise, the distributed service is compromised.

\section{Adversary model and Security Requirements}
\label{section:radis_adversary_model}
In this section, first we describe an adversary model in a distributed IoT service setting, and then we define the required security properties for a distributed remote attestation protocol.
\subsection{Adversary model}

The main goal of an adversary $Adv$ is to compromise the execution or the results of a distributed IoT service. Thus, the aim of remote attestation is to detect the distributed services which are compromised or maliciously influenced by $Adv$. We consider the following possible actions of an $Adv$ against distributed IoT services:
\begin{itemize}
\item \textbf{Software adversary.}
$Adv$ can compromise the binaries of the services, can inject malicious code in the free space of the program memory of a device, or can exploit at run-time a service
vulnerability to manipulate the data memory of a device (e.g., by corrupting control-flow pointers or data pointers).
%Additionally, \textbf{$Adv$ can forge the attestation result by managing to produce a valid result despite the software modifications present on the device.}

\item \textbf{Communication adversary.}
$Adv$ can eavesdrop on and alter the communication data between services. $Adv$ will be particularly interested to alter the communication data in such way that it will change the intended control-flow of the invoked service.

\item \textbf{Replay attack.}
$Adv$ precomputes the operations of the attestation procedure, and reports to $Vrf$ a previous valid response which hides the attack.
\end{itemize}
\textbf{Assumptions.}
Like in other attestation schemes, we rule out physical attacks, and we assume that a software adversary cannot compromise hardware-protected memory. While we do not consider Denial of Service (DoS) attacks, we limit these attacks by using a symmetric key for the service invocations, thus, a device does not perform intensive computations to refuse a fake service request. We also assume that software attacks and Man in the middle (MITM) attacks impact the control-flow of a service software.
Furthermore, we rule out an adversary that relocates itself without affecting the control-flow of the distributed services at the attestation time.
We also assume that services will respond during the attestation procedure. However, since RADIS includes the attestation result in the service invocations, typically a non-responding service will generate a timeout message, and consequently, the final attestation result will not comprise the information about the non-responsive service.

\subsection{Security requirements}
\label{sec:requirements}
In order to be resilient to the above attacks, the remote attestation scheme of distributed services should satisfy the following security properties:
\begin{itemize}
\item \textbf{Authenticity and integrity of services:} The attestation scheme should perform software integrity verification of a distributed service to guarantee that the distributed service has not been modified by any software adversary. In particular, the protocol should provide authentic and reliable evidence to prove that at run-time a distributed service has followed a legitimate control-flow. The attestation scheme should guarantee the integrity and authenticity of each of the services that compose a distributed service.

\item \textbf{Integrity of communication data:} 
The attestation scheme should detect the compromised state of distributed services when a MITM attack, which alters the communication data between two distributed services, causes the invoked service to execute a non-intended control-flow.
Each distributed service should be able to verify the trustworthy origin of its inputs, and it should reject any service calls invoked by an unauthorized device. 
\item \textbf{Freshness:} To be resilient to replay attack, any service should not be able to reply to the attestation request of $Vrf$ with a pre-computed value that could hide an ongoing attack on the service. 
Likewise, an invoked service should prove to the calling service the freshness of the response it provides to the caller.
\end{itemize}

\section{Preliminaries}
\label{section:radis_preliminaries}

In order to achieve all security properties described above, our attestation scheme requires the following components.

 \textbf{Signature scheme.}
A signature algorithm $\sigma \gets sig(sk;m)$ takes as input a message $m$ and a secret signing key $sk$ and outputs a signature $\sigma$. A verification algorithm $\{0,1\} \gets vrfsig(pk; m, \sigma)$ verifies whether $\sigma$ is valid or invalid on input of a message $m$, a signature $\sigma$, and a public verification key $pk$.

 \textbf{Message authentication code (MAC).} MAC is a pair of polynomial time algorithms $signMac()$ and $verifyMac()$ such that $\mu \gets signMac(k;m)$ outputs a MAC tag $\mu$ on input of $m$ and $k$, and $\{0,1\} \gets verifyMac(k;m, \mu)$ verifies $\mu$ on input of $m$ and $k$.
 
\begin{figure}[ht]
\centerline{\includegraphics[scale=0.9]{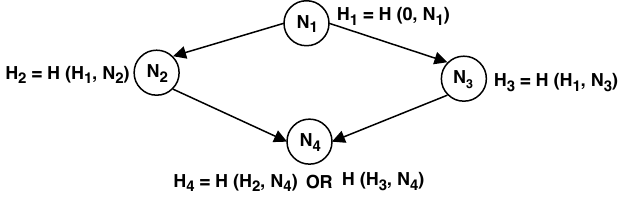}}
\caption{Hashing algorithm of Control Flow Graph}
\label{hashing_controlflow}
\end{figure}

 \textbf{Graph hashing.}
A Control Flow Graph represents the legitimate execution flows of a given software. For instance, \autoref{hashing_controlflow} depicts two valid execution flows: $N_1 \rightarrow N_2 \rightarrow N_4$ and $N_1 \rightarrow N_3 \rightarrow N_4$, where each graph node $N_1$ .. $N_4$ denotes a software instruction or a group of uninterrupted sequences of instructions, i.e., basic blocks.
We borrow the hash engine from C-FLAT, which associates each valid execution flow of a single device with a unique hash value, computing $H_{l}$ = $Hash(H_{l-1}, N_l)$. 

\section{Remote attestation of distributed IoT services: RADIS}
\label{section:radis_solution_description}
We now describe RADIS, our remote attestation protocol for distributed IoT services.

RADIS has two main operation modes: setup mode and attestation mode. Setup mode is an initial procedure, executed only once, which allows trustworthy execution of the remote attestation protocol. Attestation mode is a periodical procedure initiated by $Vrf$ at an arbitrary time.
In \autoref{tab:notation_summary}, we summarize the terms used in RADIS.
\begin{table}[ht]
\centering
\caption {Notation Summary of RADIS protocol} \label{tab:notation_summary}
\scalebox{.99}{
\begin{tabular}{llr}
\hline
Term    & Description \\
\hline  \\
$OP$      & System Operator \\
$Vrf$     & Verifier of a distributed IoT system \\
$SK_{Vrf}$    & Secret key of $Vrf$ \\
$PK_{Vrf}$    & Public key of $Vrf$ \\
$D_i$     & Device $i$ \\
$Prv_i$   & Prover $i$ \\
$sk_i$    & Secret key of $D_i$ \\
$pk_i$    & Public key of $D_i$ \\
$k_{ij}$  & shared symmetric key between $D_i$ and $D_j$ \\
$S_iu$    & Unique name of a service running on $D_i$ \\
$SFG$     & Service Flow Graph \\
%$LHV_i$   & Local Hash Value of the execution flow \\ &
%for the current service running on $D_i$ \\
$GHV_i$   & Global Hash Value stored in $D_i$ for the \\ & control-flow  execution of a service flow \\
\hline
Procedure    & Description \\ 
\hline \\
$signMac(k;m)$		& generates MAC tag on $m$\\
$verifyMac(k;m, \mu)$  & verifies MAC tag $\mu$ on $m$ using $k$\\
$sig(sk;m)$ & encrypts a message $m$ using a secret key $sk$\\
$vrfsig(pk; m, \sigma)$  & verifies $\sigma$ on $m$ using public key $pk$\\
$servatt()$      & performs attestation for a given service \\
$callatt()$      & a calling service sends an attestation \\ &request to an invoked service\\
$respatt()$      & reports attestation result from an invoked \\ & service to a calling service\\
$getrespatt()$   & retrieves the attestation response from an\\ & invoked service to a calling service\\
\hline
\end{tabular}}
\end{table}

\subsection{Setup phase}
Setup phase includes two operations: key setup and service measurement, executed respectively by $OP$ and $Vrf$.

\textbf{Key setup.}
To establish a secure communication between $Vrf$ and $Prv$, each deployed device $D_{i}$ knows $Vrf$'s public key $PK_{Vrf}$ and owns an asymmetric key-pair $(pk_{i}, sk_{i})$. In addition, two devices $D_{i}$ and $D_{j}$ that will interact in the network establish a shared symmetric attestation Message Authentication Code (MAC) key $k_{ij}$. The secret signing key $sk_{i}$ and the shared attestation key $k_{ij}$ are both stored within hardware-protected memory, preventing untrusted parties from using these keys.
%Note that the key setup process between devices is managed by $OP$, and this paper does not describe the details about the key management scheme. 
Alternatively, as a lightweight key exchanging scheme between devices can be used a random key predistribution scheme  \cite{Eschenauer:2002,Chan:2003} which rely on probabilistic key sharing among devices. The basic idea is that each device is initialized with $m$ keys, selected from a large pool of $S$ keys, such that two random subsets of size $m$ in $S$ will share at least one key with some probability $p$. Next, devices will perform shared-key-discovery to find out which of other devices they share a key with.
Note that the key setup process between devices is managed by $OP$, and this work assumes that two device $D_i$ and $D_j$ share a symmetric key $k_{ij}$ without providing details about the key management scheme.

\begin{figure}[htbp!]
\centerline{\includegraphics[scale=0.8]{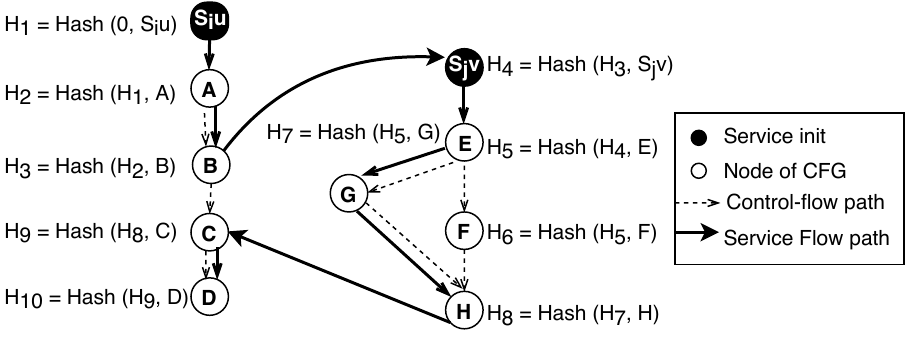}}
\caption{Hashing procedure for a legitimate Service Flow in RADIS}
\label{fig:hashingsfg}
\end{figure}

\textbf{Service measurement.}
Service measurement is a one-time-only procedure that $Vrf$ performs offline to measure the legitimate service flows of a distributed service.
Service measurement procedure follows the assumption that $Vrf$ has access to the binaries of all the services and $Vrf$ knows in advance the legitimate interactions between IoT devices. 
%First, $Vrf$ builds a Service Flow Graph (SFG) to represent a legitimate service flow in a distributed service. In a SFG, each node is a service and the edges determine the execution order of the services. Next, $Vrf$ builds the Control Flow Graph of every service that composes a service flow. 
First, $Vrf$ builds a graph, in which the nodes represent services and the edges determine the execution order of the services in a distributed service. Next, $Vrf$ builds the Control Flow Graph of every service and builds a Service Flow Graph (SFG) to represent all the possible valid transitions that a distributed service may follow at run-time. Then, starting from each valid transition, $Vrf$ executes a measurement function to associate each legitimate service flow with a single hash value as shown in \autoref{fig:hashingsfg}. Finally, $Vrf$ stores all the generated hash values in a database.
In this initial setup phase, although the measurement of the Control Flow Graph can introduce high complexity, the $Vrf$ generates the measurements offline, so the complexity of software measurement does not impact the performance of the remote attestation procedure on the device. Moreover, a typical IoT service is expected to be less complex than traditional applications, and $Vrf$ has sufficient processing resources.

\subsection{Attestation phase}

\begin{figure*}[htbp]
%\centerline{\includegraphics[width=\textwidth]{images/algorithm.pdf}}
\centerline{\includegraphics[ scale=0.74 ]{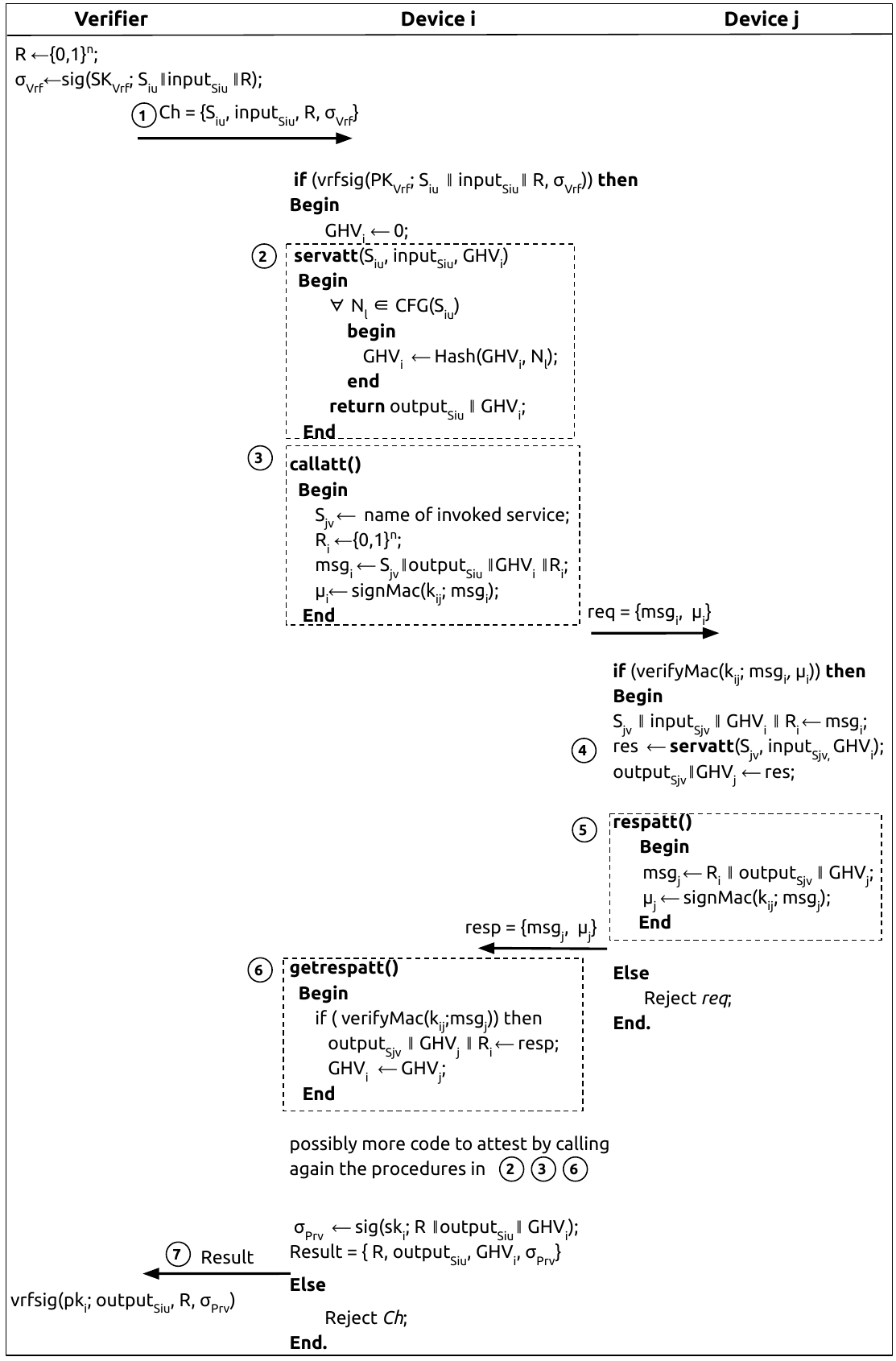}}
\caption{The algorithm of RADIS attestation protocol}
\label{fig:algorithmn}
\end{figure*}

The attestation procedure starts with $Vrf$ who sends an attestation request $Ch = {S_{iu}, input_{Siu}, R, \sigma_{Vrf}}$, where $S_{iu}$ is the name of the service to be attested, $input_{Siu}$ is the initial input for the given service $S_{iu}$, $R$ is a randomized nonce to ensure the freshness of the communication, and $\sigma_{Vrf}$ is $Vrf$'s signature over $S_{iu}$, $input_{Siu}$ and $R$ (as shown in Step \circled{\small 1} in \autoref{fig:algorithmn}). Upon receiving the attestation request $Ch$, the device $D_{i}$, which serves as a prover $Prv_i$, verifies the signature by using the $Vrf$'s public key $PK_{Vrf}$. If the signature is valid, RADIS protocol, which is running on $Prv_i$, invokes the procedure $servatt$ (Step \circled{\small 2}) to attest $S_{iu}$ with the provided input $input_{Siu}$. Since $S_{iu}$ is the first service of the service flow, $GHV_{i}$ will be initialized with 0. The invocation of $servatt$ triggers the tracing of the execution flow of $S_{iu}$, to compute a hash value for each instruction, and to store the accumulated hash value in $GHV_{i}$. 

When $S_{iu}$ invokes another service $S_{jv}$, the code of $S_{iu}$ that handles the service invocation will be attested by the procedure $callatt$ (Step \circled{\small 3}). Among the arguments of the service call, the service invocation will also include the attestation result of $S_{iu}$ and a nonce $R_i$ to initiate the attestation for $S_{jv}$. Specifically, to initiate the request, $callatt$ computes a MAC signature $\mu_i = signMac(k_{ij}; msg_i)$ over the message $msg_i$ = $S_{jv} \parallel output_{Siu} \parallel GHV_i \parallel R_{i}$, where $S_{jv}$ is the name of the invoked service, $output_{Siu}$ is the output $S_{iu}$ which serves as input data in the service call, $GHV_i$ is the attestation result of $S_{iu}$, and $R_i$ is a randomized nonce. On receiving the service request, $D_j$ uses $k_{ij}$ to verify the MAC signature $verifyMac(k_{ij}; msg_i, \mu_i)$ and prove the authenticity and integrity of the request. 
In case the service call is valid, RADIS protocol running on $D_j$ starts the attestations for $S_{jv}$ by calling $servatt$ (Step \circled{\small 4}) on the received input data. The code of $S_{jv}$ which handles the response will be attested by $respatt$ (Step \circled{\small 5}). 
Next, $D_i$  handles the response of $D_j$ by calling $getrespatt$ (Step \circled{\small 6}) and updates $GHV_i$ with the hash value $GHV_j$ produced by $D_j$. After the response, in case $S_{iu}$ continues code execution or invokes other services, RADIS will trigger again $servatt$, $callatt$, and $getrespatt$.

Upon a complete execution of all the service that compose a service flow, $Prv_{i}$ retrieves $GHV_i$ stored locally, and it sends back to $Vrf$ the signed attestation result $\sigma_{Prv}$ =$sig(sk_i;R\parallel output_{Siu}\parallel GHV_i)$ (Step \circled{\small 7}). $Vrf$ verifies the signature of the response $vrfsig(pk_i;R\parallel output_{Siu} \parallel GHV_i\parallel \sigma_{Prv}))$ and then proceeds with hash validation. Since $Vrf$ has initially stored the valid hash for each service, to validate the attestation response, $Vrf$ checks in the database whether $GHV_i$ is among the legitimate hash values saved in the database. If it matches, then $GHV_i$ serves as an evidence to prove that each service of the service flow is legitimate.

\section{Experimental setup and evaluation}
\label{section:radis_evaluation}
This section describes our experiments and presents the performance evaluation of RADIS. 

Recall from \autoref{fig:algorithmn} that RADIS protocol computes a hash value for every running service in a distributed system, and it is composed of two main computations: (1) the attestation of each individual service that composes a distributed service (performed by $servatt()$) and (2) the service request invocation and the reply obtained along with each remote service attestation (performed by $callatt()$, $respatt()$, $getrespatt()$). 
The attestation for each individual service is performed based on the control-flow of the service.
As the complexity of this computation is similar to the protocol described in \cite{Abera2016}, the complexity of the hash computation for the individual attestation is linear to the number of control-flow instructions that the service has to execute. Considering that in RADIS, the hash computation for each service starts either from an initial service (from 0) or from a previous calculated hash (as described in \autoref{section:radis_solution_description}), RADIS does not introduce additional overhead with respect to the work \cite{Abera2016} to compute the hash of each individual service.

However, in order to transmit the attestation result among services, RADIS sends a hash value in every service call in addition to the standard parameters. Due to the communication of the hash value, RADIS introduces an additional overhead compared to the service calls where no attestation  of distributed services is performed. Considering that RADIS computes a single hash over a previous calculated hash (as described in \autoref{fig:algorithmn}, procedure \textit{servatt()}), the hash length remains constant despite the number of services that can compose a distributed service. In the following, we describe the experiment and the evaluation of the additional overhead that RADIS introduces.

%Thus, RADIS introduces an additional overhead compared to the service calls where no attestation of distributed services is performed.
%Considering that in RADIS, the hash computation for each service starts either from an initial service (from 0) or from a previous calculated hash (as described in \autoref{sec:solution_description}), RADIS does not introduce additional overhead with respect to the work \cite{Abera2016} to compute the hash of each individual service.

%However, in a distributed service setting, in order to transmit the attestation result among services, RADIS invokes a hash value in every service call. Due to the communication of the hash value, RADIS introduces an additional overhead compared to the service calls where no attestation  of distributed services is performed. Despite the number of services that can compose a distributed service, the hash length remains constant. In the following, we describe the experiment and the evaluation of the additional overhead that RADIS introduces.

\subsection{Experimental Setup} 
To attest individual services, 
%we first built the Control Flow Graph (CFG) from each service source code. 
we developed a hash module and customized a trace module\footnote{We customized \textbf{trace} which is an open-source python module https://docs.python.org/2/library/trace.html.} to trace the control-flow at run-time. During execution, the customized trace module invokes the hash module to compute and accumulate a single hash value for each executed control-flow. We assume that an adversary will not be able to disable or modify the trace module and the hash module. For a secure deployment of the protocol on real devices, trace module and hash module can run within a lightweight hardware-assisted secure environment based on ARM TrustZone.

In order to measure the overhead for transmitting a hash value in every service call, we implemented a distributed service scenario composed of three services: $captureImage() \rightarrow checkImage() \rightarrow unlockDoor()$. %where an Outdoor Camera captures an image by performing $captureImage()$, a Security Monitor uses $checkImage()$ to analyze the captured image, and based on the command received from the Security Monitor, a Smart Door will run $unlockDoor()$ , as described in \autoref{sec:problem_setting}.
We implemented each service in Python v3.6.3 using Python Flask v1.0.2. We deployed each service inside a Docker container with 1GB RAM and 1.2GHz CPU running on Alpine Linux v3.8 and establish a HTTP communication among the services. We use SHA-1 and SHA-384 as a cryptographic hash function and a Keyed-Hash Message Authentication Code (HMAC) based on SHA-256 as a MAC in order to show the complexity and computational overhead of the implemented distributed service.

\subsection{Evaluation} 
For single service attestation, the overhead to compute a hash for the entire control-flow  of a service with 10 lines of code is $\approx 36$ microseconds.
We evaluated the communication overhead of RADIS by measuring the run-time of distributed services without performing the attestation protocol and with performing the attestation with the two cryptographic hash functions, namely, SHA-1 and SHA-384. 

\begin{table}[h]
\centering
\caption {RADIS run-time in seconds (s)}
\label{tab:runtime_summary} 
\begin{tabular}{lllr}
\hline
\small{Services} &\small{No attest} & \small{SHA-1} & \small{SHA-384} \\\hline  \\
\small{$captureImage-checkImage$} &0.00383s &0.01164s &0.01213s\\
\small{$checkImage-unlockDoor$} &0.00441s &0.01211s &0.01298s\\
\small{$captureImage-checkImage-unlockDoor$} &0.00750s &0.02355s &0.02503s\\
\hline
\end{tabular}
\end{table}
 
\begin{figure}[htbp]
\centerline{\includegraphics[scale=0.5]{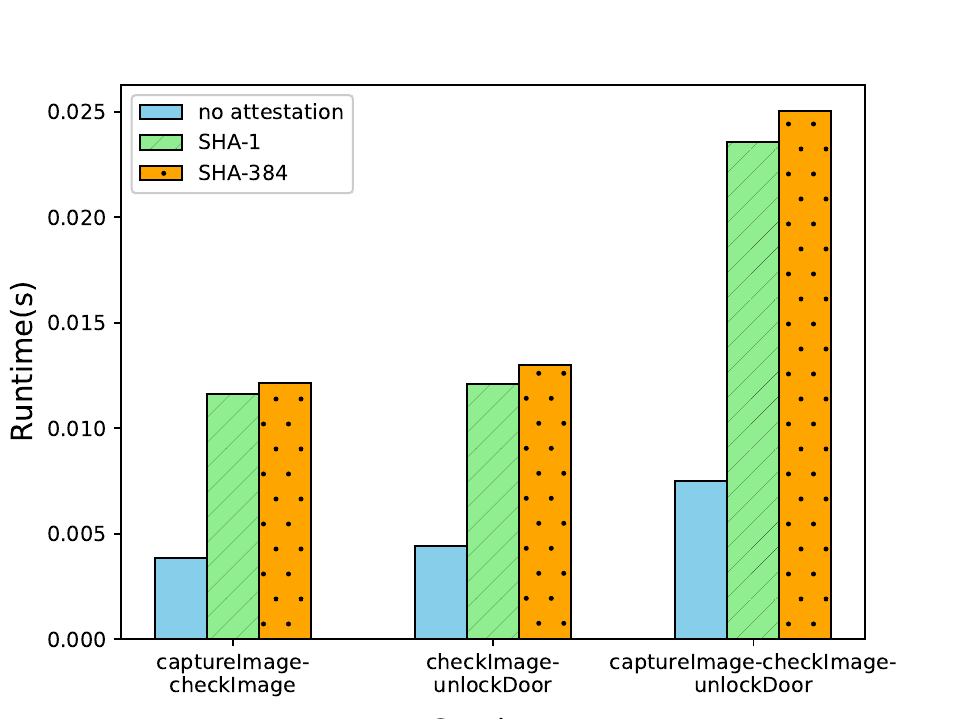}}
\caption{Comparison of RADIS performance for SHA-1 and SHA-384 for two and three services in a distributed service}
\label{fig:radis_evaluation}
\end{figure}

 From \autoref{tab:runtime_summary} one can see that the communication overhead of SHA-1 and SHA-384 among two services is respectively $\approx$ 8 milliseconds and $\approx$ 9 milliseconds with respect to the case of no attestation. While in the case of three services, the communication overhead is $\approx$ 16ms for SHA-1 and $\approx$ 17.5ms for SHA-384. The time of signature verification HMAC SHA-256 of each service is $\approx$ 1 millisecond, and it is included in the measured run-time shown in \autoref{tab:runtime_summary}. The runtime measurements of RADIS are also shown in \autoref{fig:radis_evaluation} which illustrates a comparison of RADIS performance for SHA-1 and SHA-384 for the services that compose the distributed service of our case study application. 
 
\begin{figure}[htbp]
\centerline{\includegraphics[scale=0.5]{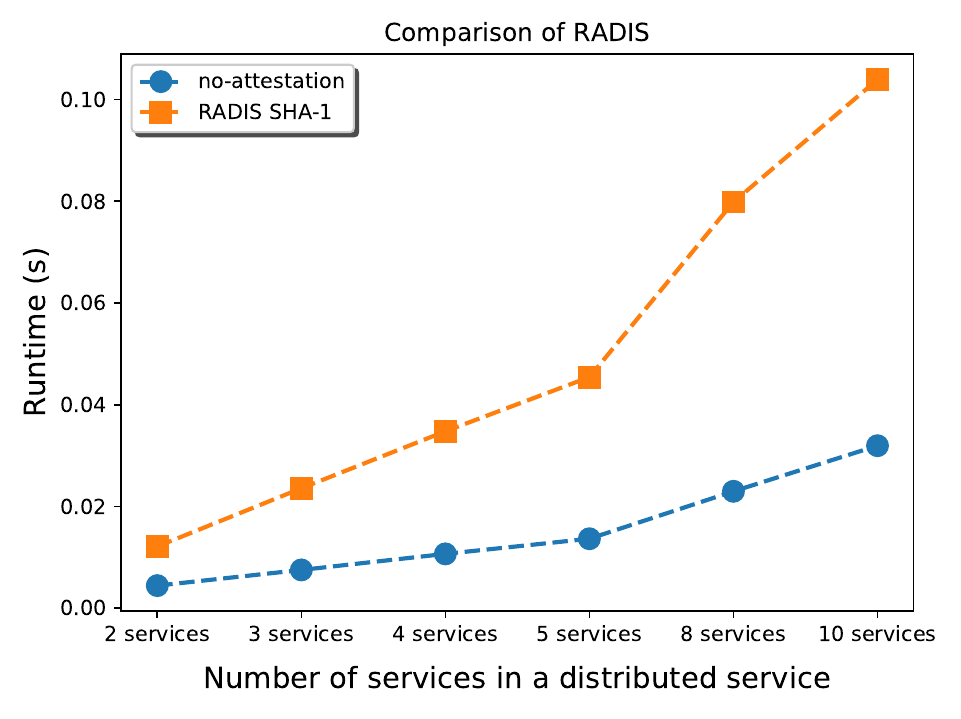}}
\caption{RADIS performance in various number of services in a distributed system}
\label{fig:compare}
\end{figure}

 Our experiments show that the communication overhead  between two services is constant. Therefore, for a distributed service which comprises N services, the overhead is linear in (N-1). Let $T_{noattest}$ be the time of interaction between services when no attestation is performed and $T_{overhead}$ is the overhead of RADIS between two services. The runtime $T_{RADIS}$ of the communication between N services in RADIS can be given as $T_{RADIS} = T_{noattest} + T_{overhead} * (N-1)$. The scalability of RADIS for N services depends on scalability properties of the underlying architecture of a distributed service. See \autoref{fig:compare} that reports the overhead of RADIS for SHA-1 in a various number of services that compose a distributed system.
 %In our experiments, the time of signature verification HMAC SHA-256 of each service is $\approx$ 1 millisecond, and it is included in the measured run-time shown in \autoref{tab:runtime_summary}. 
 The results confirm that the performance of RADIS is reasonable for attesting distributed IoT services.

\vspace{-.1em}
\section{Security Analysis}
\label{section:radis_security_analysis}

This section presents some arguments to give an insight into the proof that RADIS meets the security requirements described in \autoref{sec:requirements}.

\textbf{Authenticity and Integrity of software:} In RADIS, a trace module intercepts the control-flow of each service at run-time and invokes a hash module to compute a  cumulative hash. Thus, $Adv$ will not be able to execute an arbitrary code or change the control-flow that will not be observed by the hash module. Following the assumptions that the hash functions are collision-resistant, and that $Adv$ cannot disable or modify the code of the trace and hash module, then $Adv$ will not be able to generate a valid hash value for an altered control-flow. Additionally, RADIS intercepts the service calls, and each invoked service first registers the attestation result of the calling service, and then starts the execution. Hence, a data attack on the calling service that produces a corrupted output which changes the control-flow of the invoked service will produce an unknown hash value to $Vrf$. As only RADIS can access the secret signing key $sk$, the final attestation result %reported to $Vrf$ 
is authenticated and cannot be tampered by $Adv$.

\textbf{Integrity of communication:} 
Any changes of the communication data by $Adv$ that effects the execution flow of the invoked service will produce an unknown hash value to $Vrf$. The communication data between two devices is authenticated with a MAC symmetric encryption $k_{ij}$. Given a secure MAC function, it will be infeasible for $Adv$ to forge the data without knowing $k_{ij}$.

 \textbf{Freshness:} The freshness of the attestation is ensured by a randomized nonce $R$ sent by $Vrf$, and randomized nonces $R_i$ exchanged among device $D_i$. Assuming that the probability of sending a randomized nonce $R$, where $R=R_{old}$ is negligible, two different attestation results will not match. Therefore, $Vrf$ can detect the replay attack.
 
 \section{Conclusions and Future works}
\label{section:radis_conclusions}
While IoT systems become interoperable, an important challenge for the remote attestation schemes is to guarantee the trustworthy state of the IoT services that compose a distributed service. A secure interaction between devices is a key issue in IoT systems, and in this paper, we emphasize the need for a distributed services attestation in IoT systems. We presented RADIS, as a protocol that provides a comprehensive and reliable integrity check of a distributed service. Our solution gives evidence about the trustworthiness of the services that compose a distributed services and the interaction flow between services. 

Future work includes the optimization of hash computation for resource-constrained devices and implementation of our protocol on real devices. In addition, we plan to investigate lightweight authentication schemes for distributed IoT services. Finally, we are looking for possible solutions for development of low cost lightweight easy to deploy remote attestation scheme for dynamic IoT swarms.

 \section{Acknowledgement}
\label{sec:acknowlege}
This work has been partially funded by the Progetto Ateneo Sapienza 2019, "PRIvacy-preserving, Security, and MAchine-learning techniques for healthcare applications (PRISMA)".
\bibliographystyle{splncs04}
\bibliography{main.bib}

\begin{thebibliography}{10}
\providecommand{\url}[1]{\texttt{#1}}
\providecommand{\urlprefix}{URL }
\providecommand{\doi}[1]{https://doi.org/#1}

\bibitem{Abera2016}
Abera, T., Asokan, N., Davi, L., Ekberg, J.E., Nyman, T., Paverd, A., Sadeghi,
  A.R., Tsudik, G.: C-{FLAT}: {C}ontrol-{F}low {A}ttestation for {E}mbedded
  {S}ystems {S}oftware. In: Proceedings of the 2016 {ACM} {SIGSAC} Conference
  on Computer and Communications Security {CCS '16} (2016)

\bibitem{PADS}
{Ambrosin}, M., {Conti}, M., {Lazzeretti}, R., {Masoom Rabbani}, M., {Ranise},
  S.: {PADS: Practical Attestation for Highly Dynamic Swarm Topologies}. ArXiv
  e-prints  (2018)

\bibitem{Ambrosin2016}
Ambrosin, M., Conti, M., Ibrahim, A., Neven, G., Sadeghi, A.R., Schunter, M.:
  {SANA}: {S}ecure and {S}calable {A}ggregate {N}etwork {A}ttestation. In:
  Proceedings of the 2016 {ACM} {SIGSAC} Conference on Computer and
  Communications Security {CCS} '16 (2016)

\bibitem{Asokan2015}
Asokan, N., Brasser, F., Ibrahim, A., Sadeghi, A.R., Schunter, M., Tsudik, G.,
  Wachsmann, C.: {SEDA}: {Scalable Embedded Device Attestation}. In:
  Proceedings of the 22nd {ACM} {SIGSAC} Conference on Computer and
  Communications Security {CCS '15} (2015)

\bibitem{Chan:2003}
Chan, H., Perrig, A., Song, D.: Random key predistribution schemes for sensor
  networks. In: Proceedings of the 2003 IEEE Symposium on Security and Privacy.
  SP '03 (2003)

\bibitem{Dessouky:2018}
Dessouky, G., Abera, T., Ibrahim, A., Sadeghi, A.R.: Litehax: lightweight
  hardware-assisted attestation of program execution. In: 2018 IEEE/ACM
  International Conference on Computer-Aided Design (ICCAD). pp.~1--8 (2018)

\bibitem{Dessouky2017}
Dessouky, G., Zeitouni, S., Nyman, T., Paverd, A., Davi, L., Koeberl, P.,
  Asokan, N., Sadeghi, A.R.: {LO}-{FAT}: {Low-Overhead} {Control} {Flow}
  {ATtestation} in {Hardware}. In: Proceedings of the 54th Annual Design
  Automation Conference 2017 {DAC} '17 (2017)

\bibitem{Eschenauer:2002}
Eschenauer, L., Gligor, V.D.: A key-management scheme for distributed sensor
  networks. In: Proceedings of the 9th ACM Conference on Computer and
  Communications Security {CCS '02} (2002)

\bibitem{Gu2009}
Gu, L., Cheng, Y., Ding, X., Deng, R.H., Guo, Y., Shao, W.: {Remote Attestation
  on Function Execution (Work-in-progress)}. In: Proceedings of the First
  International Conference on Trusted Systems on - {INTRUST} '09 (2010)

\bibitem{DOP}
{Hu}, H., {Shinde}, S., {Adrian}, S., {Chua}, Z.L., {Saxena}, P., {Liang}, Z.:
  {Data-Oriented Programming: On the Expressiveness of Non-control Data
  Attacks}. In: {2016 IEEE Symposium on Security and Privacy SP '16} (2016)

\bibitem{Ibrahim:2016}
Ibrahim, A., Sadeghi, A.R., Tsudik, G., Zeitouni, S.: {DARPA: Device
  attestation resilient to physical attacks}. In: Proceedings of the 9th ACM
  Conference on Security \& Privacy in Wireless and Mobile Networks {WiSec '16}
  (2016)

\bibitem{Kil2009}
Kil, C., Sezer, E.C., Azab, A.M., Ning, P., Zhang, X.: Remote attestation to
  dynamic system properties: Towards providing complete system integrity
  evidence. In: 2009 {IEEE}/{IFIP} International Conference on Dependable
  Systems {\&} Networks (2009)

\bibitem{Kohnhauser:2017}
Kohnh{\"a}user, F., B{\"u}scher, N., Gabmeyer, S., Katzenbeisser, S.: {SCAPI: a
  scalable attestation protocol to detect software and physical attacks}. In:
  Proceedings of the 10th ACM Conference on Security and Privacy in Wireless
  and Mobile Networks {WiSec '17} (2017)

\bibitem{Kohnhauser:2018}
Kohnh\"{a}user, F., B\"{u}scher, N., Katzenbeisser, S.: {SALAD: Secure and
  Lightweight Attestation of Highly Dynamic and Disruptive Networks}. In:
  Proceedings of the 2018 on Asia Conference on Computer and Communications
  Security {ASIACCS '18} (2018)

\bibitem{mirai}
KrebsonSecurity: Mirai botnet.
  \url{http://krebsonsecurity.com/tag/mirai-botnet} (October 2016), [Online;
  accessed 15-September-2018]

\bibitem{Ronen2017}
Ronen, E., Shamir, A., Weingarten, A.O., OFlynn, C.: {IoT} goes nuclear:
  Creating a {ZigBee} chain reaction. In: 2017 {IEEE} Symposium on Security and
  Privacy ({SP}) (2017)

\bibitem{senrio}
Senrio: Devil's ivy: Flaw in widely used third-party code impacts millions.
  \url{http://blog.senr.io/blog/devils-ivy-flaw-in-widely-used-third-party-code-impacts-millions}
  (July 2017), [Online; accessed 15-September-2018]

\bibitem{Shacham2007}
Shacham, H.: The geometry of innocent flesh on the bone. In: Proceedings of the
  14th {ACM} conference on Computer and communications security {CCS} '07
  (2007)

\bibitem{Shi}
Shi, E., Perrig, A., Doorn, L.V.: {BIND}: A {F}ine-{G}rained {A}ttestation
  {S}ervice for {S}ecure {D}istributed {S}ystems. In: 2005 {IEEE} Symposium on
  Security and Privacy {(SP)} (2005)

\bibitem{Zeitouni2017}
Zeitouni, S., Dessouky, G., Arias, O., Sullivan, D., Ibrahim, A., Jin, Y.,
  Sadeghi, A.R.: {ATRIUM: Runtime attestation resilient under memory attacks}.
  In: 2017 IEEE/ACM International Conference on Computer-Aided Design (ICCAD).
  pp. 384--391 (2017)

\end{thebibliography}
\end{document}